\documentstyle[aps,epsfig,multicol,rotate]{revtex}
\def\beginwide{
        \end{multicols} \vspace*{-0.5cm} \noindent
        \rule{3.5in}{.1mm}\rule{.1mm}{5mm} \widetext \medskip }
\def\beginwidetop{
        \end{multicols} \vspace*{-0.5cm} \noindent
        \widetext \medskip }
\def\endwide{
        \hspace*{3.35in}~\rule[-5mm]{.1mm}{5mm}\rule{3.5in}{.1mm}
        \begin{multicols}{2} \vspace*{-1.0cm} \noindent }
\def\endwidebottom{
        \begin{multicols}{2} \vspace*{-1.0cm} \noindent }

\begin{document}

\title{An extended-phase-space dynamics for the generalized nonextensive 
thermostatistics} 
 
\author{J. S. Andrade Jr., M. P. Almeida, A. A. Moreira, and G. A. Farias}

\address{
Departamento de F\'\i sica, Universidade Federal do Cear\'a,\\
Campus do Pici, Caixa Postal 6030, 60455-760 Fortaleza, Cear\'a, Brazil
}
\date{\today}
\maketitle 
\draft
\begin{abstract}
We apply a variant of the Nos\'e-Hoover thermostat to derive the
Hamiltonian of a nonextensive system that is compatible with the
canonical ensemble of the generalized thermostatistics of
Tsallis. This microdynamical approach provides a deterministic
connection between the generalized nonextensive entropy and power law
behavior. For the case of a simple one-dimensional harmonic
oscillator, we confirm by numerical simulation of the dynamics that
the distribution of energy $H$ follows precisely the canonical
$q$-statistics for different values of the parameter $q$. The approach
is further tested for classical many-particle systems by means of
molecular dynamics simulations. The results indicate that the
intrinsic nonlinear features of the nonextensive formalism are capable
to generate energy fluctuations that obey anomalous probability
laws. For $q<1$ a broad distribution of energy is observed, while for
$q>1$ the resulting distribution is confined to a compact support.
\end{abstract} 
\begin{multicols}{2}

Since the pioneering work of Tsallis in 1988 \cite{Tsallis88} where a
nonextensive generalization of the Boltzmann-Gibbs (BG) formalism for
statistical mechanics has been proposed, intensive research
\cite{www} has been dedicated to develop the conceptual framework
behind this new thermodynamical approach and to apply it to realistic
physical systems. In order to justify the generalization of Tsallis,
it has been frequently argued that the BG statistical mechanics has a
domain of applicability restricted to systems with short-range
interactions and non (multi)fractal boundary conditions
\cite{Tsallis99}. Moreover, it has been recalled that anomalies
displayed by mesoscopic dissipative systems and strongly non-Markovian
processes represent clear evidence of the departure from BG
thermostatistics. These types of arguments have been duly reinforced
by recent convincing examples of physical systems that are far better
described in terms of the generalized formalism than in the usual
context of the BG thermodynamics (see \cite{Tsallis99} and references
therein).  It thus became evident that the intrinsic nonlinear
features present in the Tsallis formalism that lead naturally to power
laws represent powerful ingredients for the description of complex
systems.

In the majority of studies dealing with the thermostatistics of
Tsallis, the starting point is the expression for the generalized
entropy $S_q$,
\begin{equation}\label{Sq}
S_q=\frac{k}{q-1}\left[1-\int{[f(x)]^{q}dx}\right]~,
\end{equation}
where $k$ is a positive constant, $q$ a parameter and $f$ is the
probability distribution. Under a different framework, some
interesting studies \cite{Lyra97} have shown that the parameter $q$ can
be somehow linked to the system sensibility on initial conditions. Few
works have been committed to substantiate the form of entropy
(\ref{Sq}) in physical systems based entirely on first principles
\cite{Abe00,Plastino94}. For example, it has been demonstrated
that it is possible to develop dynamical thermostat schemes which are
compatible with the generalized canonical ensemble
\cite{Plastino97}. In a recent study by one of us \cite{Murilo01}, a
derivation of the generalized canonical distribution is presented from
first principle statistical mechanics. As a consequence, it is shown
that the particular features of a macroscopic subunit of the canonical
system, namely, the heat bath, determines the nonextensive signature
of its thermostatistics and therefore its power law behavior. More
precisely, it is exactly demonstrated in \cite{Murilo01} that if one
specifies the capacity of the heat bath as
\begin{equation}\label{dbeta}
\frac{dE}{d(1/\beta)}=\frac{1}{q-1}~,
\end{equation}
where $q$ is a constant, $1/\beta \propto kT$, and $T$ is the
temperature, the generalized canonical distribution maximizing
(\ref{Sq}) for the Hamiltonian $H$ of the system is recovered,
\begin{equation}\label{canon}
f(H) \propto [1+\beta (q-1)(E-H)]^{\frac{1}{q-1}}~.
\end{equation}
Here, $E$ is a conserved quantity and denotes the energy of the
extended system (system+heat bath). Equation~(\ref{canon}) provides a
very simple but meaningful connection between the generalized
$q$-statistics and the thermodynamics of nonextensive systems. It is
analogous to state that, if the condition of an infinite heat bath
capacity is violated, the resulting canonical distribution can no
longer be of the exponential form and therefore should not follow the
traditional BG thermostatistics. In the present letter, we will show
how the conjecture proposed in \cite{Murilo01} can be used to develop a
variant of the original Nos\'e-Hoover thermostat \cite{Nose84,Hoover85} 
that is consistent with the $q$-thermostatistics. We will then
validate the technique by applying it to the cases of a simple
harmonic oscillator and a classical many-particle system.

We consider a system of $N$ particles having coordinates $x_{i}'$,
masses $m_i$ and potential energy $\Phi(x')$. As in the extended system
method originally proposed by Nos\'e, here we also introduce an
additional degree of freedom through a variable $s$ which will play
the role of an external heat bath, acting as to keep the average of
the kinetic energy at a constant value. In practice, this is achieved
by simply rescaling the {\it real} variables in terms of a new set of 
{\it virtual} variables
\begin{equation}\label{scale}
{\bf x}^{\prime}_i = {\bf x}_i~,~~
{\bf p}^{\prime}_i = \frac{{\bf p}_{i}}{s^\lambda}~,~~ 
p^{\prime}_s = \frac{p_s}{s^\lambda}~,~~
t^{\prime} = \int{\frac{dt}{s}}~,
\end{equation}
where $\lambda$ is a rescaling exponent, and 
$({\bf x}_{i}^{\prime},{\bf p}_{i}^{\prime},t^{\prime})$ and 
$({\bf x}_{i},{\bf p}_{i},t)$ are the real and virtual coordinates, momenta
and time, respectively. At this point, we postulate that a generalized
Hamiltonian for the extended system can be written as
\begin{equation}\label{hamvir}
H_{q}({\bf x} , {{\bf p}} , p_s , s)\!=\!\sum_{i=1}\! 
{ {\bf p}_{i}^2 \over2m_is^{2\lambda}}
 \!+\! \Phi ({\bf x})\!+\!{p_{s}^2 \over {2Q}}
 \!+\! {1 \over \alpha}{{{s^\gamma}\!-\!1} \over {\gamma}},
\end{equation}
where the first two terms on the right side represents the energy of
the physical system that is free to fluctuate \cite{Nosefoot}. The
virtual variable $p_{s}$ also has a real counterpart,
$p^{\prime}_s=p_s/s^\lambda$, and has been introduced to allow for a
dynamical description of the variable $s$. More precisely, the third
term $p_{s}^2/{2Q}$ corresponds to the kinetic energy of the heat bath
and the parameter $Q$ is an inertial factor associated with the motion
of the variable $s$. The last term of the Hamiltonian (\ref{hamvir})
is a power law potential in $s$. As we show next, it provides the
essential link between the concept of extended-phase-space dynamics
and the generalized canonical ensemble.

We start by considering the quasi-ergodic hypothesis and writing the time 
average of a given quantity $A({\bf x}^{\prime},{\bf p}^{\prime})$ as 
\begin{eqnarray}\label{average}
\overline{A}
={1\over Z}\int \int \int \int d{\bf x} \, d{\bf p} \, dp_{s} \,ds \, 
{A \over s } \delta (H_{q}-E) \\
\text{with}~~~~~
Z=\int\int\int\int d{\bf x}\,d{\bf p}\,dp_{s}\,ds\,s^{-1}\delta(H_{q}-E)~,
\nonumber
\end{eqnarray}
where $Z$ is the analogous to a microcanonical normalization factor
for the generalized Hamiltonian (\ref{hamvir}). The factor $s^{-1}$ in
(\ref{average}) accounts for the variability in the real time sampling
during the extended-phase-space dynamics \cite{Nose84}. Transforming the
virtual momenta ${\bf p}$ and coordinates ${\bf x}$ back to real
variables, changing the order of integration and rewriting the volume
element as $d{\bf x}\,d{\bf p}=s^{g\lambda} d{\bf x}^{\prime} d{\bf
p}^{\prime}$, where $g$ is the number of degrees of freedom, we obtain
\begin{equation}\label{inter1}
\overline{A}\!
= \! {1\over Z} \! \int \!\! \int \!  d{\bf x}^{\prime} d{\bf p}^{\prime} A
\! \int \!\! \int dp_{s}\,ds\,s^{g\lambda -1} \delta (H_{q}\!-\!E)~.
\end{equation}
If we now make use of the property of the $\delta$ function, $\delta
[h(s)]=\delta (s-s_0) / h^{\prime}(s_0)$, where $s_0$ is the zero of
$h$, it follows that
\begin{eqnarray}
\overline{A}
&=&{1\over Z} \int \int d{\bf x}^{\prime} \, d{\bf p}^{\prime} \, A \nonumber \\ 
&\times& \int dp_s \, \alpha
\left[ 1 + \alpha \gamma \left(E-H-{{p_s}^2 \over {2Q}}\right) \right] 
^{{g\lambda \over \gamma} -1}~, 
\end{eqnarray}
where,
\begin{equation}
H=H( {\bf x}^{\prime}, {\bf p}^{\prime}) = 
\sum_{i=1} {{\bf p}^{\prime 2}_{i} \over2m_i} + \Phi ({\bf x}^{\prime})~.
\end{equation}
By integration with respect to $p_s$ we get 
\begin{eqnarray}
\overline{A}	
&=&{1\over Z} 
{\left({\alpha Q\over 2 \gamma} \right)^{1/2}}
B\left({1 \over 2},{g\lambda \over \gamma}\right)\nonumber \\
&\times& \int \int d{\bf x}^{\prime} \, d{\bf p}^{\prime} \, A 
{[1 + \alpha \gamma (E-H)]}^{{g\lambda \over \gamma} -{1 \over2}}~, 
\end{eqnarray}
where $B$ is the beta function, $B(z,w)=\int_{0}^{1}t^{z-1}(1-t)^{w-1}dt$.
Finally, if we define $\alpha={\beta(q+1)/2g\lambda}$ and 
$\gamma = {{2g\lambda(q-1)}/({q+1})}$, the generalized canonical average 
is recovered
\begin{eqnarray}
\overline{A}\!	
&=&\!{1\over Z^{\prime}}\! \int\! \int \! d{\bf x}^{\prime} 
d{\bf p}^{\prime} A
\left[1\! + \!\beta (q\!-\!1){\left(E\!-\!H\right)} \right]
^{1 \over (q\!-\!1)}\\ 
\text{with}&&
Z^{\prime}\!=\!\int\!\int d{\bf x}^{\prime} d{\bf p}^{\prime} 
\left[1\! + \!\beta (q\!-\!1){\left(E\!-\!H\right)} \right]
^{1 \over (q\!-\!1)}~, \nonumber
\end{eqnarray}
and we have thus proved that, under conservation of the extended
Hamiltonian Eq.~(\ref{hamvir}), the fluctuations in the energy
$H({\bf x}^{\prime}, {\bf p}^{\prime})$ of the physical
system should be consistent with the canonical formulation of the
nonextensive $q$-thermostatistics.

It is possible to confirm the validity of this approach with
a simple realization of the generalized thermostat scheme. We
consider an extended system composed of a single one-dimensional
harmonic oscillator coupled to a heat bath whose thermal capacity
obeys essentially Eq.~(\ref{dbeta}). From Eq.~(\ref{hamvir}),
such a system can be described by the following extended Hamiltonian:
\begin{equation}\label{winkler}
H_{q}(x,p,p_s,s) = {p^2 \over2s^{2\lambda}}+{x^2 \over2} + {p_{s}^2 \over {2Q}}
  + {1 \over{\alpha}} {s^{\gamma} -1 \over \gamma}~.
\end{equation}
Here $m=1$ for simplicity and we choose to set $\lambda=2$ because 
the nonlinear dynamics for this case when $q=1$ (i.e., for the BG
thermostatistics) has been shown to be sufficiently chaotic
to generate average properties of the canonical ensemble
\cite{Winkler92}. From (\ref{winkler}) and the scaling relations
(\ref{scale}), we obtain the equations of motion for 
the extended system in the {\it real} phase space,
\begin{eqnarray}
{dx^{\prime}\over dt^{\prime}}&=&{p^{\prime}\over s} \nonumber \\
{dp^{\prime}\over dt^{\prime}}&=&-{x^{\prime}\over s} 
- {2s^2p^{\prime}_s p^{\prime}\over Q}\nonumber \\
{ds^{\prime}\over dt^{\prime}}&=&{s^3p^{\prime}_{s} \over Q} \nonumber \\
{dp^{\prime}_{s}\over dt^{\prime}}&=&{1 \over s^2} \left( 2p^{\prime 2} 
- {1\over \alpha}s^{\gamma} \right)
-{2s^2 p^{\prime 2} _s \over Q}~.
\end{eqnarray}
A fifth order Runge-Kutta subroutine is then used to numerically solve
this set of nonlinear differential equations. To ensure the
conservation of energy $H$ and the stability of integration, all runs
have been performed with $10^8$ time steps of $\Delta
t^{\prime}=10^{-4}$ each. The density maps shown in Figs.~1(a) and (b)
for $q=0.8$ and $1.2$, respectively, provide clear evidence that the
dynamics of both systems fills space. For $q>1$ [see Fig.~1(b)], the
accessible phase-space lies in a compact set, whereas the phase-space
support for $q<1$ [see Fig.~1(a)] is infinite. The former situation
is compatible with the necessary cut-off condition on energy for $q>1$
\cite{Tsallis99}. In Fig.~2 we show the logarithmic plot of
the distributions of the transformed variable
$\chi=1+\beta(q-1)(E-H)$, where $H=p^{\prime 2}/2+{x}^{\prime 2}/2$,
for three different values of the parameter $\gamma=4(q-1)/(q+1)$
corresponding to $q=0.7$, $0.8$ and $0.9$. Indeed, we observe in all
cases that the fluctuations in $\chi$ follow very closely the
prescribed power law behavior, $\rho(\chi)\propto \chi^{1/(q-1)}$, and
therefore confirm the validity of our dynamical approach to the
generalized canonical ensemble. As shown in Fig.~3, the simulations
performed for $q>1$ are also compatible with the expected scaling
behavior. However, instead of the long-range tail obtained for the
case $q<1$, a rather unusual power law with positive exponent is
observed.

Now we focus on a more complex application of the thermostat
scheme introduced here. The basic idea is to simulate, through
molecular dynamics (MD), the nonextensive behavior of a classical
many-particle system. For completeness, we start by rewriting the
expression for the extended Hamiltonian (\ref{hamvir}) in terms of the
usual q-thermostatistics parameters (i.e., $q$ and $\beta$)
\begin{equation}\label{hamtsa}
H_{q}({\bf x} , {\bf p} , p_s , s)\! =\! \sum_{i=1} \!{ p_{i}^2 \over2m_i
s^{2\lambda}}
 \! + \!\Phi ({\bf x}) \!+ \!{p_{s}^2 \over {2Q}}
 \! + \!{1 \over{\beta}} ln_{q} \left(s^{{2g\lambda}\over{q +1}}\right), 
\end{equation}
where $ln_{q}(s)\equiv (s^{q-1}-1)/(q-1)$ \cite{Tsallis99}. From
Eq.~(\ref{hamtsa}), it is then possible to derive the equations of
motion for any value of $q$ and any type of effective potential of
interaction. We consider a cell containing $108$ identical particles that
interact through the Lennard--Jones potential, $\Phi(\Delta x_{ij})=4
\epsilon [(\sigma /\Delta x_{ij})^{12}-(\sigma /\Delta x_{ij})^{6}]$,
where $\Delta x_{ij}$ is the distance between particles $i$ and $j$,
$\epsilon$ is the minimum energy and $\sigma$ the zero of the
potential. Distance, energy and time are measured in units of
$\sigma$, $\epsilon$ and $(m \sigma)^2 /{\epsilon}$, respectively, and
the equations of motion are numerically integrated using a
predictor-corrector algorithm \cite{Allen87}.

Compared to the previous example of a single harmonic oscillator, the
complexity of the many-particle system hinders a quantitative
prediction of the statistical behavior of its energy fluctuations.
Because the exact form or even a plausible approximation of the
density of states $\Omega (H)$ is difficult to obtain in this case, we
restrict ourselves to the qualitative analysis of the resulting energy
distribution $\rho(H)\propto \Omega(H) f(H)$. Furthermore, we
performed additional simulation tests with different number of
particles and physical conditions to confirm that the MD system is
always led to unstable trajectories in phase-space whenever the value
of $q$ is set to be smaller than a given threshold $q_{\rm min}$. In
spite of these limitations, however, the results shown in Fig.~4
clearly indicate the tendency for a broader distribution of energy
when $q<1$ (we set $q$ to be slightly larger than $q_{\rm min}\approx
0.9940$ in this case). For $q>1$, on the other hand, the resulting
distribution of energy is notably more confined than the Maxwellian
distribution obtained for $q=1$ (see Fig.~4).

In summary, we have shown that the essential features of the
generalized canonical distribution can be captured with a proper
extension of the standard Nos\'e-Hoover thermostat. To the best of our
knowledge, this is the first time that a Hamiltonian approach to the
nonextensive $q$-thermostatistics leads explicitly to the observation
of a power law behavior ($q<1$). We thus believe that the
microdynamical formalism presented in this work can provide a
deterministic link between the generalized entropy conjecture
Eq.~(\ref{Sq}) and the concept of L\'evy flights
\cite{Plastino97,Alemany94}. Finally, the methodology introduced here
is flexible enough to accommodate the description of other
nonextensive systems of physical significance.

This work has been supported by the Brazilian National Research 
Council CNPq.

\begin{figure}
\epsfig{file=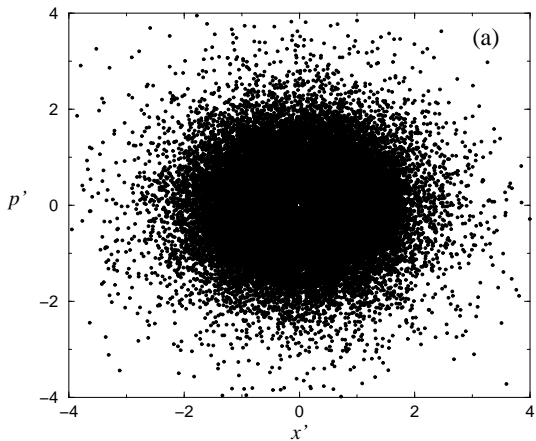,width=7cm}
\caption{
(a) Density plot of the harmonic oscillator dynamics subjected to 
the generalized thermostat scheme for $q=0.8$. The initial conditions
are [$x^{\prime}(0)=0.5$, $p^{\prime}(0)=0.5$, $s(0)=1.0$, $p_s^{\prime}(0)=0.0$]
and the thermostat parameters have been set to $\alpha=1.0$ and $Q=1.0$.
(b) Same as (a) but for $q=1.2$.
}
\end{figure}

\begin{figure}
\epsfig{file=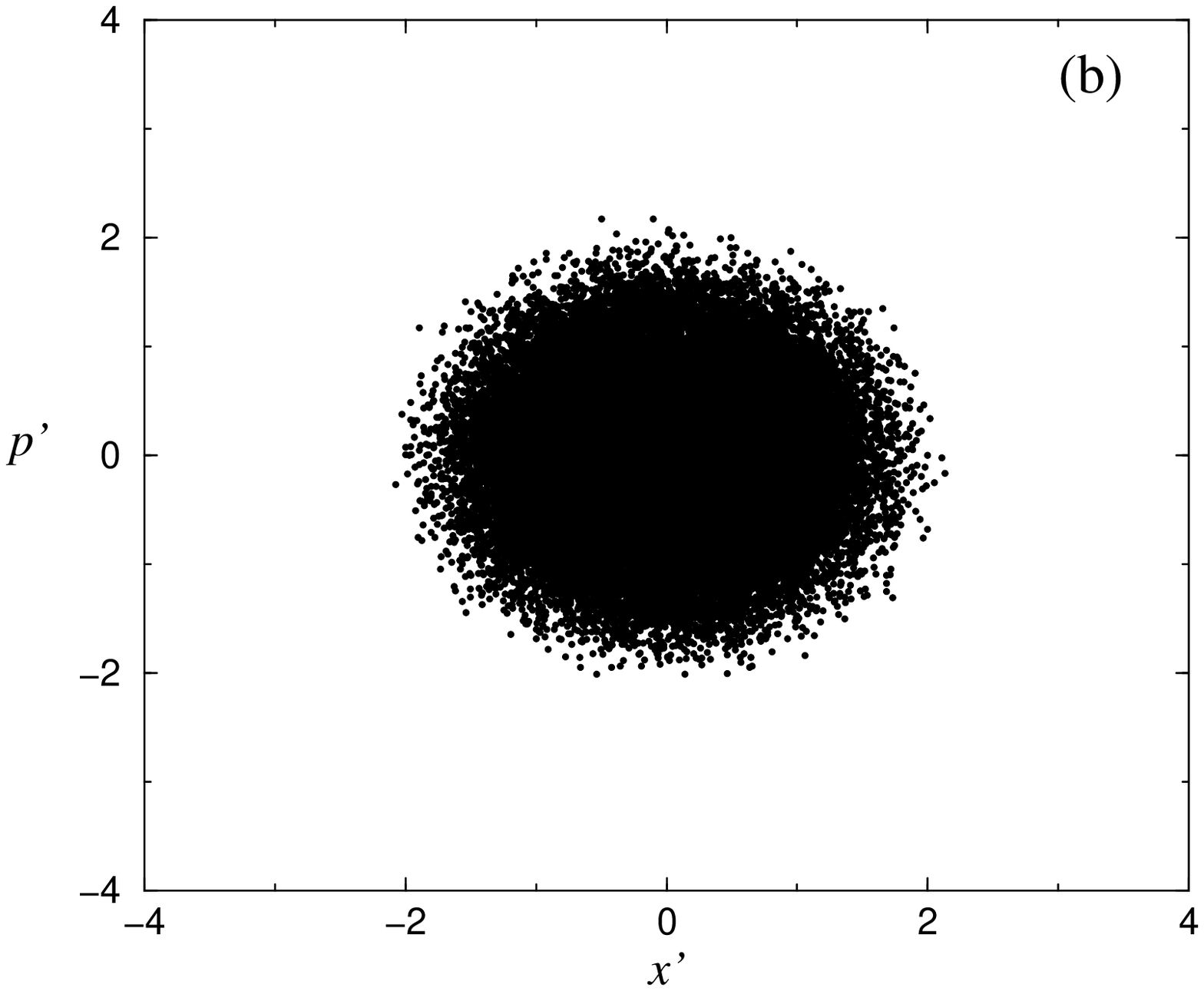,width=7cm}
\end{figure}

\begin{figure}
\epsfig{file=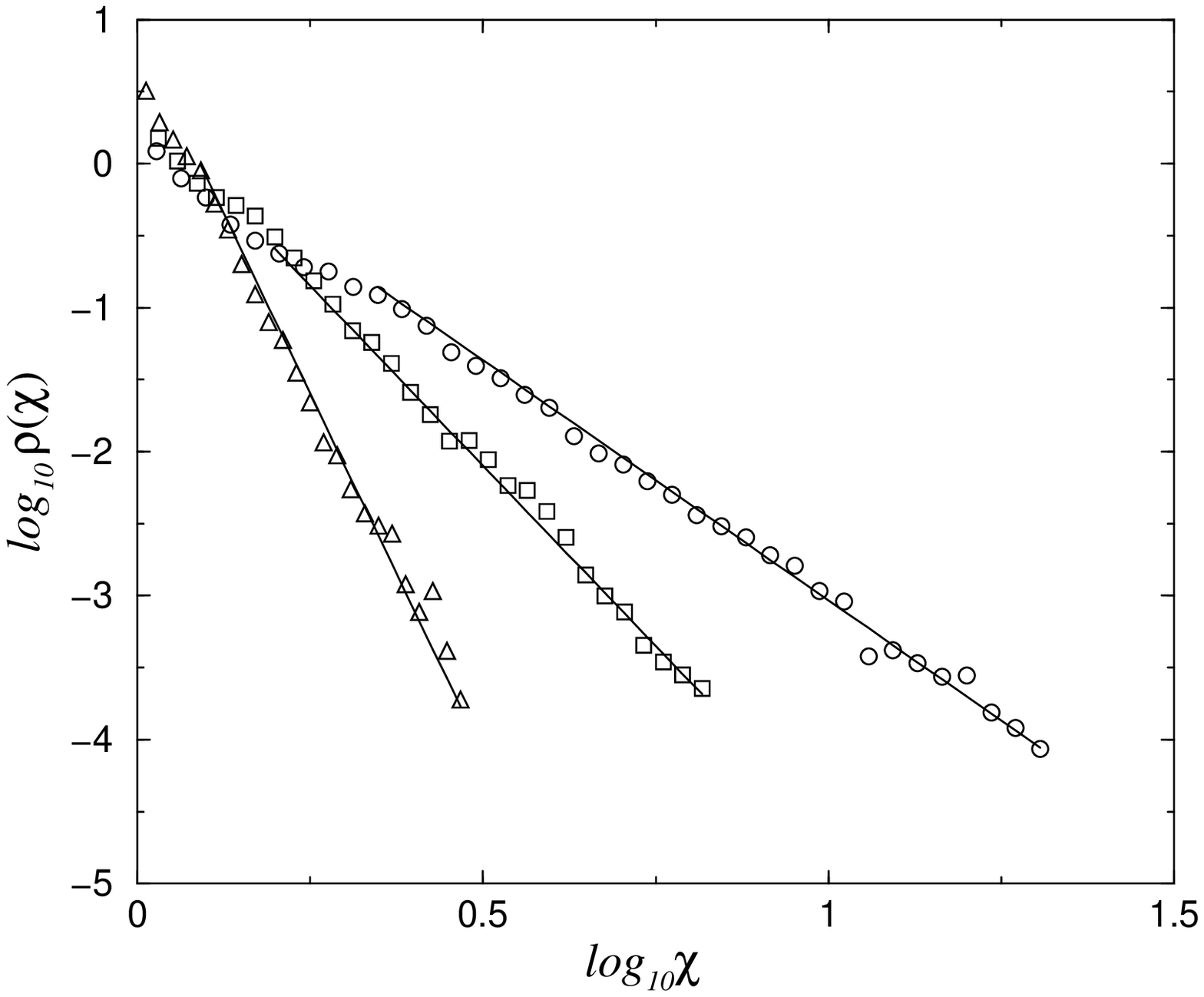,width=7cm}
\caption{
Logarithmic plot of the distributions of the transformed variable
$\chi$ for $q=0.7$ (circles), $0.8$ (squares) and $0.9$ (triangles). 
From right to left, the three straight lines with slopes $-3.33$, 
$-5.0$ and $-10.0$ correspond to the expected power law behavior
$\rho(\chi)\propto \chi^{1/(q-1)}$.
}
\end{figure}

\begin{figure}
\epsfig{file=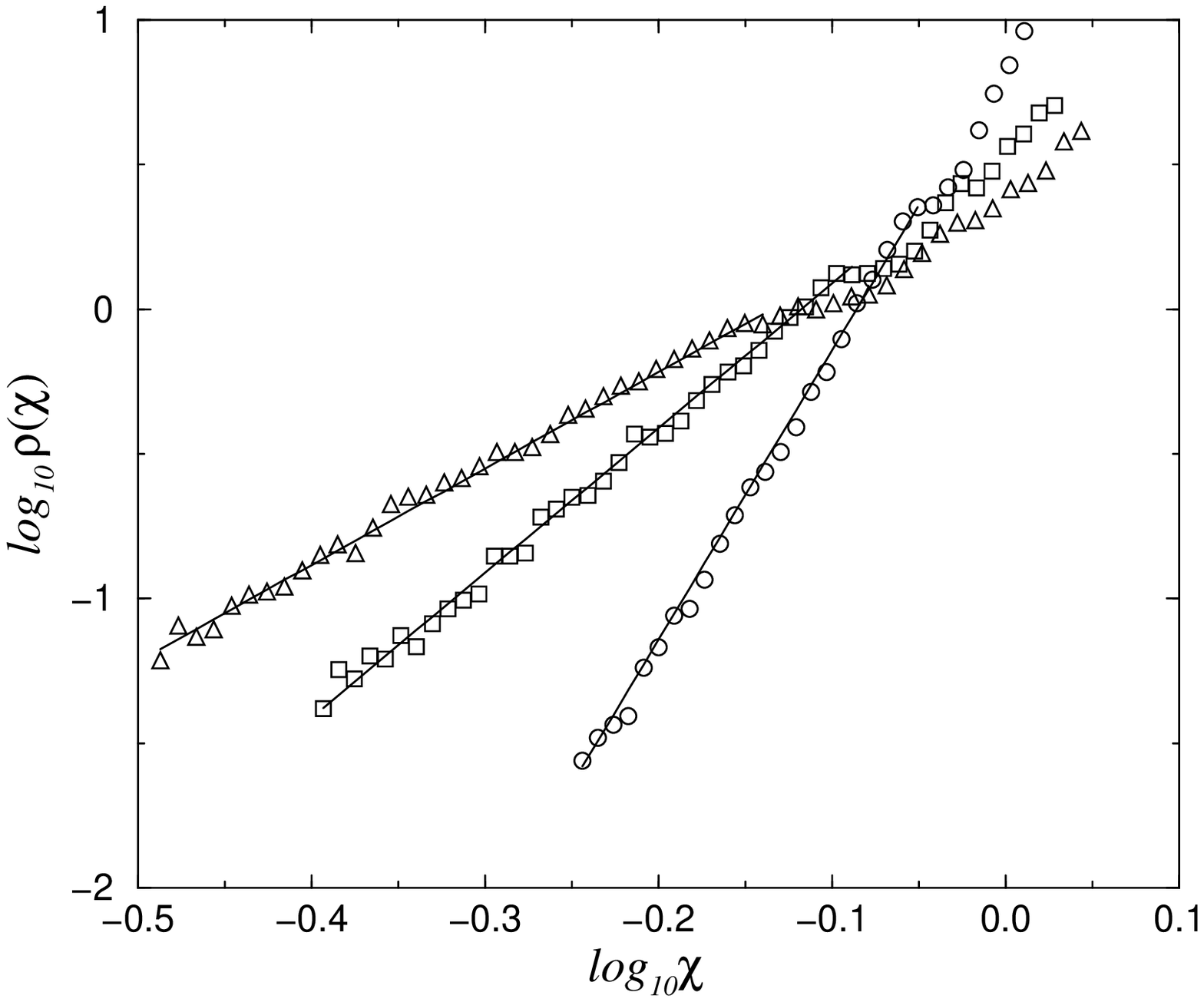,width=7cm}
\caption{
Logarithmic plot of the distributions of the transformed variable
$\chi$ for $q=1.1$ (circles), $1.2$ (squares) and $1.3$ (triangles). 
From right to left, the three straight lines with slopes $10.0$, 
$5.0$ and $3.33$ correspond to the expected power law behavior
$\rho(\chi)\propto \chi^{1/(q-1)}$.
}
\end{figure}

\begin{figure}
\epsfig{file=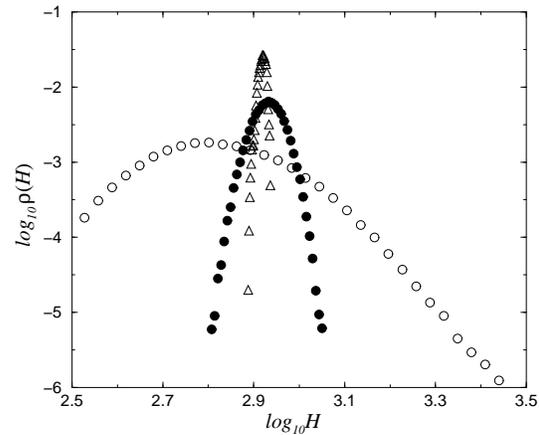,width=7cm}
\caption{
Logarithmic plot of the energy distributions for $q=0.9941$ (circles), 
$1.0$ (full circles) and $1.1$ (triangles). In all three cases, the MD
simulations have been performed with 108 particles, $\beta=0.2$, and 
a density of $0.1$ particles/$\sigma^3$.
}
\end{figure}

\end{multicols}


\begin{thebibliography}{99}
\bibitem{Nosefoot} A similar Hamiltonian expression has been previously 
used in \cite{Nose84} as a counter-example to demonstrate that $\ln (s)$ is 
the only possible potential form leading to the BG canonical distribution.
\end{thebibliography}
\end{document}